\documentclass[review]{elsarticle}

\usepackage{hyperref}
\usepackage{graphics} 
\usepackage{epsfig} 
\usepackage{mathptmx} 
\usepackage{times} 
\usepackage{amsmath} 
\usepackage{amssymb}  
\usepackage{subcaption}
\usepackage{adjustbox}
\usepackage{siunitx}
\usepackage{graphicx}
\usepackage{booktabs}
\usepackage{multirow}
\usepackage{soul}
\usepackage{url}
\usepackage[dvipsnames]{xcolor}
\usepackage{color}
\usepackage[labelformat=parens,labelsep=quad, skip=3pt]{caption}
\journal{arXiv}

\begin{document}

\begin{frontmatter}





\title{\LARGE \bf
Confidence Aware Neural Networks for Skin Cancer Detection}






\author[add1]{Donya Khaledyan}
\author[add2]{AmirReza Tajally}
\author[add3]{Ali Sarkhosh}
\author[add4]{Afshar Shamsi}
\author[add4]{Hamzeh Asgharnezhad \corref{mycorrespondingauthor}}
\cortext[mycorrespondingauthor]{Corresponding author}
\ead{hamzeh.asgharnezhad@gmail.com}
\author[add5]{Abbas Khosravi, Senior, IEEE}
\author[add5]{Saeid Nahavandi, Fellow, IEEE}

\address[add1]{Faculty of Electrical Engineering, Beheshti University, Tehran, Iran}
\address[add2]{Department of Industrial Engineering, University of Tehran, Tehran, Iran}
\address[add3]{Department of Electrical and Computer, Isfahan University of Technology (IUT), Isfahan, Iran}
\address[add4]{Individual Researcher, Tehran, Iran}
\address[add5]{A. Khosravi, and S. Nahavandi are with the
Institute for Intelligent Systems Research and Innovation (IISRI), Deakin University, Australia}

\begin{abstract}
Deep learning (DL) models have received particular attention in medical imaging due to their promising pattern recognition capabilities. However, Deep Neural Networks (DNNs) require a huge amount of data, and because of the lack of sufficient data in this field, transfer learning can be a great solution. DNNs used for disease diagnosis meticulously concentrate on improving the accuracy of predictions without providing a figure about their confidence of predictions. Knowing how much a DNN model is confident in a computer-aided diagnosis model is necessary for gaining clinicians’ confidence and trust in DL-based solutions. To address this issue,  this work presents three different methods for quantifying uncertainties for skin cancer detection from images.
It also comprehensively evaluates and compares performance of these DNNs using novel uncertainty-related metrics. The obtained results reveal that the predictive uncertainty estimation methods are capable of flagging risky and erroneous predictions with a high uncertainty estimate. We also demonstrate that ensemble approaches are more reliable in capturing uncertainties through inference.
\end{abstract}


\end{frontmatter}

\section{INTRODUCTION}

Skin cancer is mainly caused due to the unusual growth of skin cells. It often develops when the body is unprotected from the sunlight. Skin cancer detection in the early stages is a problematic even for dermatologists. As with any disease, the early detection of skin cancer could lead to much better treatment results.\\
To diagnose skin cancer, various parameters must be considered, including asymmetry of the lesion, color, border, diameter, and the development of the lesion are the parameters that should be considered. There are three major types of skin cancer: basal-cell skin cancer (BCC), squamous-cell skin cancer (SCC), and melanoma~\cite{diepgen2002epidemiology}. The first two are less common, but melanoma is the deadliest type of skin cancer and one of the rare types of this disease. Melanoma is caused by the uncontrolled growth of a kind of skin cell called melanocyte. There are Many factors contribute to skin cancer including family history of skin cancer, UV radiation, ionizing radiation, smoking, photosensitizing drugs~\cite{stratton2019feasibility}. From the therapeutic point of view, skin cancer is one of the most expensive types of cancer. Its early diagnosis can make the condition better as with Melanoma there is approximately 95\% treatment rate if it is diagnosed and treated in initial stages~\cite{lorenzo2019clinical}.\\
Generally, computer-aided diagnosis can be a useful tool, mainly in areas where there is a shortage of experts~\cite{heidari2019development,heidari2018applying,rafieipour2020study}.
Several traditional statistical works are presented for skin cancer prediction, but in recent time, we are witnessing the emergence of deep learning and specifically the convolution neural networks~\cite{mashhadi2020deep,heidari2021applying,altaher3using}. Machine learning is an updated version of traditional statistical methods and accomplishes high-dimensional information process contributing to disease diagnosis in the field of computer-aided diagnosis~\cite{heidari2020new,khuzani2021applying, AfsharUncertainty}. It has ben shown that CNNs match or even outperform medical experts for detecting skin cancer from images. \\
In the field of medical diagnoses, the protestation against infrequent but costly faults is necessary. Deep learning models usually have a shortage demonstration of uncertainty and offer overoptimistic and overconfident predictions. Presenting uncertainty models is essential for decision-making choices~\cite{szatmari2019comparison}.
Measurement uncertainty is significant to risk estimation and decision-making. For example, organizations make decisions every day based on quantitative measurement data reports. If estimation results are not valid, then the risk of the decision will increase. Choosing the wrong plans could affect business goals in that organization. This case is an example in a million examples in our world that shows how measurement results impact decisions. If the capacity to evaluate the quality of the measurement outcomes were present, individuals and organizations could decide more confidently. For delicate applications, such as medical image analysis, and self-driving cars, the neural network must provide uncertainty predictions for its decisions. More precisely, the network representing when it is probable to be incorrect or uncertain ~\cite{rahmati2019predicting}. \\
Monte-Carlo dropout (MCD) is broadly prevalent for its straightforwardness and effectiveness among the numerous choices of uncertainty quantification in deep neural networks.
Gal and Ghahramani in~\cite{gal2016dropout} introduced an astonishingly straightforward technique for measuring model uncertainty. They discovered that training any NNs with dropouts, classically used for stopping overfitting, can be understood as an estimated inference of the posterior weight, given that dropouts are added after every weighted layer.\\
In recent years, the idea of ensemble results in the designing and proposing of breakthrough algorithms.
Ensemble networks are also known as an alternative technique for quantifying uncertainty. In~\cite{lakshminarayanan2016simple}, they scaled up the idea of ensembling for capturing uncertainty related to the model. In comparison to the other training methods, ensemble learning is more robust and accurate on classification and prediction~\cite{karri2017transfer}.\\
A deep neural network can have thousands of parameters; therefore, training a new model from scratch is time-consuming and computationally expensive, especially when access to a robust dataset is impossible. In this situation, transfer learning can propose a pre-built model to be fine-tuned to a new problem and its dataset. This procedure saves the time to train a network from scratch and has been instrumental in making deep learning accessible to non-expert users~\cite{cheplygina2019not}.\\
In this paper, we use three algorithms for DNN uncertainty quantification: MCD, Ensemble Bayesian network, and Ensemble Monte Carlo Dropout (EMCD). The latest on is a combination of MCD and ensemble, for quantifying the uncertainty related to model (epistemic). It should be noted that EMCD is a quite new method that uses the advantages of both the MCD technique and ensemble network. We use novel performance metrics for the quantitative and comprehensive evaluation of uncertainty predictions~\cite{asgharnezhad2020objective}. The uncertainty prediction evaluation is estimated in a similar way to that of binary classification evaluation. Through experimentation, we  clarify whether aforementioned methods  can generate high uncertainty for erroneous predictions. This research is done both qualitatively (visually) and quantitatively (using new uncertainty evaluation criteria). The whole setting will give additional information to assist physicians productively. \\
The rest of the paper is organized in the following way. Section \ref{Sec:Dataset} defines the specification of the dataset. Section \ref{Sec:Related} describes the related works and the background of our work. The proposed algorithms for uncertainty prediction of skin cancer are introduced in section~\ref{Sec:Method}. Section~\ref{Sec:Uncertainty quantification} describes the quantitative metrics used for uncertainty approximation. Section~\ref{Sec:Experimental} defines the detail of the proposed networks. Section~\ref{Sec:Results} discusses the achieved results and simulations in detail. Finally, the study is summarized with conclusion and the future work in section~\ref{Sec:Concl}.


\section{Dataset}\label{Sec:Dataset}

The paper uses MNIST HAM-10000 dataset for conducting experiments. This dataset includes dermoscopy images which are available on Kaggle~\cite{kaggle}.
HAM10000 (Human Against Machine with 10000 training images) includes 10015 images of cancer and non-cancer cases. At least 50\% of lesions in this dataset have been confirmed by pathology, while the rest of the cases were approved by expert consensus or validated by in-vivo confocal microscopy.
These images are from diverse populations with different age ranges. The number of images in this dataset is satisfactory to be used for other goals, including image segmentation, classification, feature extraction, and transfer learning. \\


\begin{figure}
	\centering
	\begin{subfigure}{0.44\linewidth}
		\includegraphics[width=\linewidth]{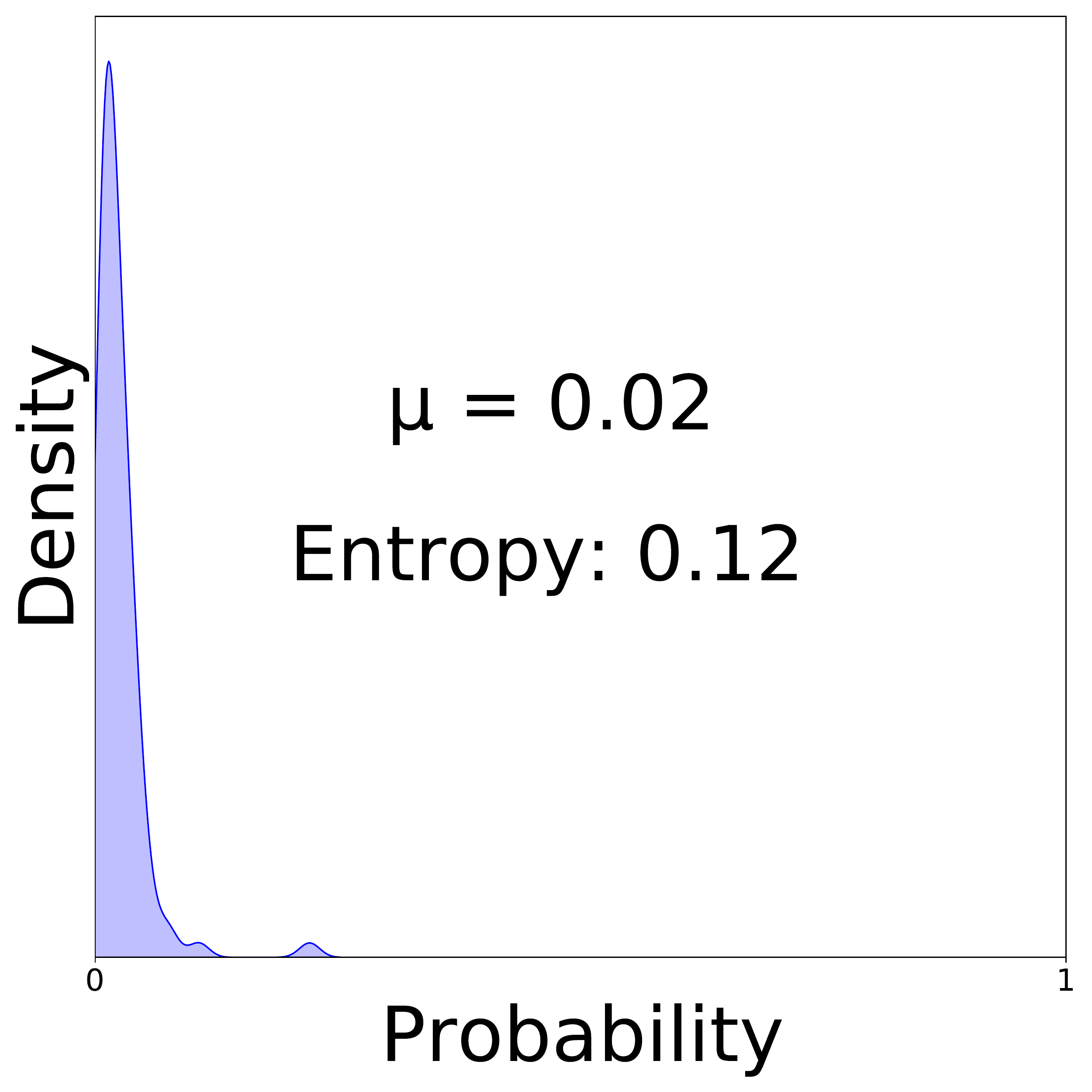}
		\caption{Healthy and certain}
		\label{fig: Certain}
		
     \end{subfigure}
     \begin{subfigure}{0.44\linewidth}
	  \centering
		\includegraphics[width = \linewidth]{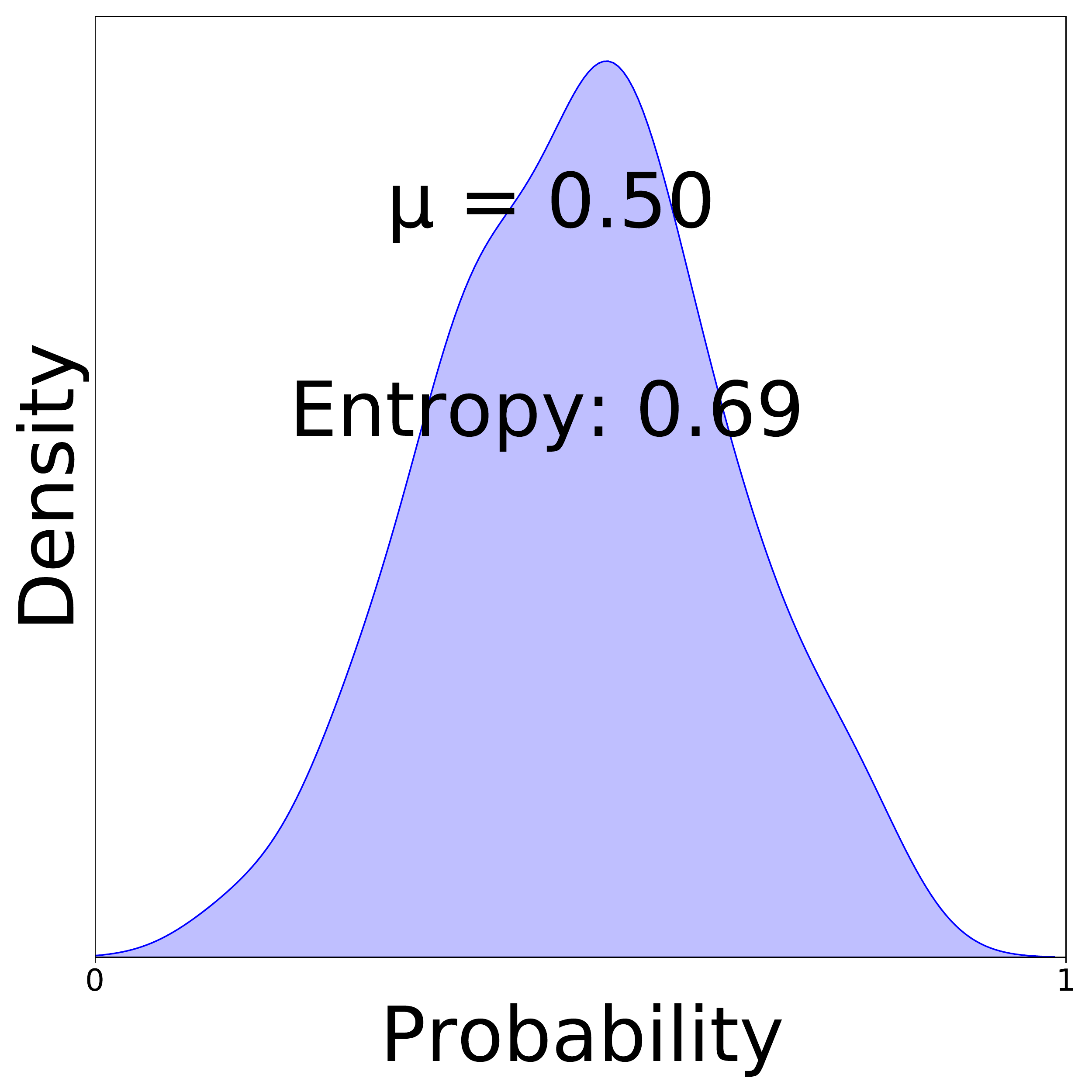}
		\caption{Healthy and uncertain}
		\label{fig: Uncertain}
		
		\end{subfigure}
	\caption{The figure represents the posterior distributions of two normal images.}
	\label{Fig: Images}
\end{figure}

\begin{figure}[t]
\centering
\includegraphics[width=0.88\textwidth]{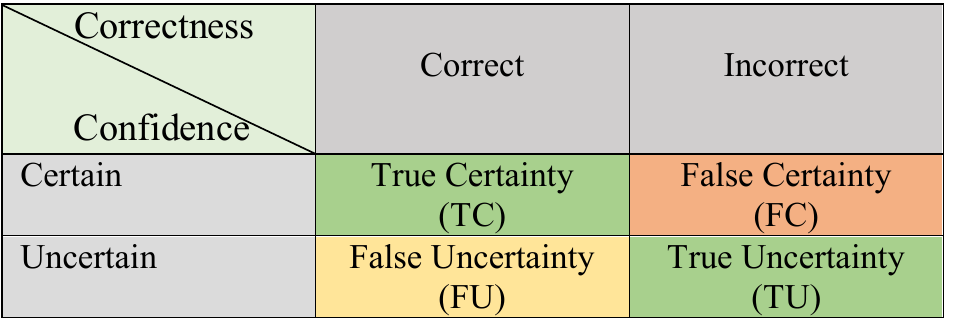}
\caption{The confusion matrix of uncertainty is represented in this figure}
\label{Fig: Confusion matrix}
\end{figure}


\begin{figure}
	\centering
	
		\includegraphics[width=\linewidth]{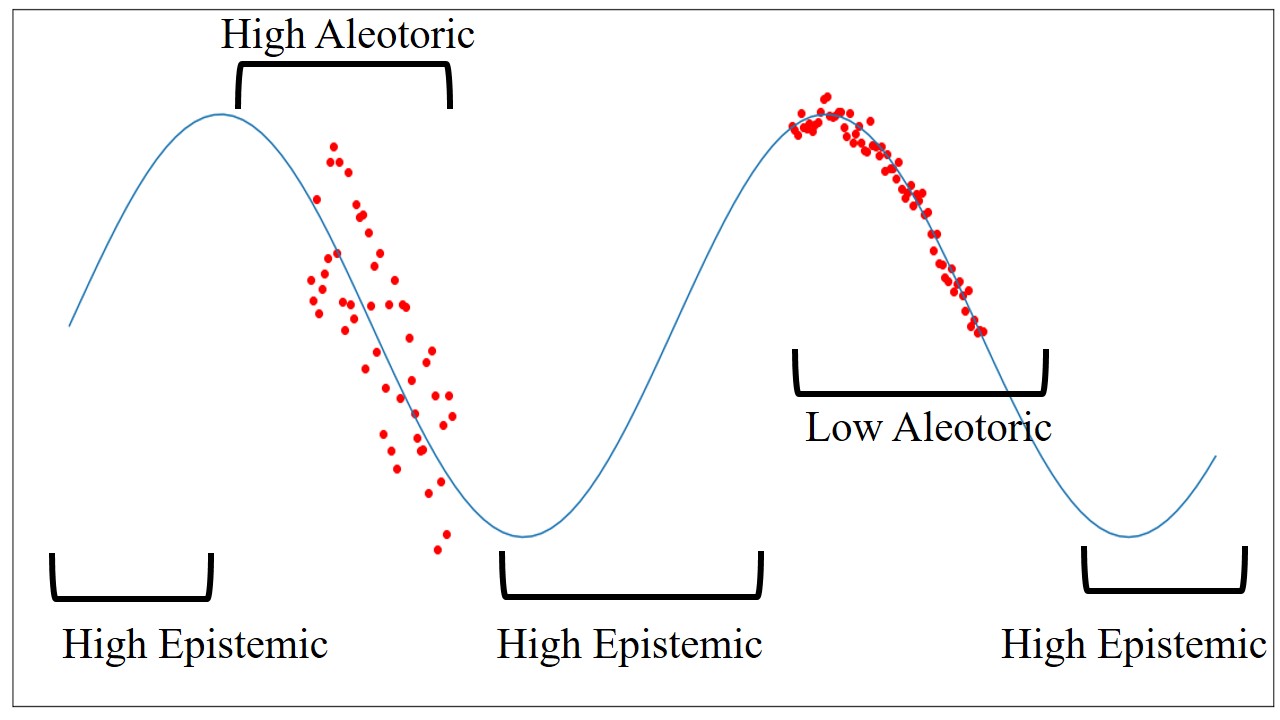}
		\caption{ A demonstration of the different types of uncertainty in a sinus regression context}
		\label{fig: slide}
		
     \end{figure}



\section{Related Work}\label{Sec:Related}
CNNs have been commonly used for the medical image processing and image classification.
Authors in~\cite{brinker2018skin} proposed the first organized and systematic research on the classification of skin lesion diseases. They mainly concentrated on the application of CNNs for classification of skin cancer. The authors also discussed the challenges which must be addressed in the classification task.
Han et al. in~\cite{han2018classification} performed a classifier for 12 different skin abnormalities and diseases based on a clinical database. They used a fine-tuned ResNet with the 1938 Asan dataset, atlas site images, and MED-NODE dataset. Images used in this research are all limited to a specific age range.
Haenssle et al.~\cite{haenssle2018man} proposed the first work, which compares the functionality of CNN with a worldwide group of 58 dermatologists for the skin cancer classification. Most of the dermatologists were outperformed by CNN. Furthermore, the authors concluded that, dermatologists can benefit from CNN assistance image classification regardless of their training and expereince. In this work, Google's Inception v4 CNN architecture was trained and validated using a dermoscopic dataset and corresponding diagnoses. Authors in~\cite{marchetti2018results} presented a cross-sectional work by using 100 randomly chosen dermoscopic cases that included 50 melanomas, 6 lentigines, and 44 nevi. This paper was proposed in an international computer vision melanoma challenge dataset. The authors established a combination of 5 techniques for the classification task.
Authors in~\cite{khan2021attributes} proposed a detection and recognition system for skin lesions. The prime goal of this work is the melanoma cancer detection. The proposed method is based on deep transfer learning for both the segmentation step (MASK-RCNN) and classification (DenseNet). They reached an average accuracy of 96.3\% on ISBI2016, 94.8\% on ISBI2017, and 88.5\% on HAM 10000 datasets.\\
As an alternative tool, uncertainty prediction can provide extra information to the experts, especially in the medical field, because sometimes a correct or incorrect diagnosis can be worth human life. Through gaining a more general view of the network, the trust in the output will also increase, and if necessary, extra experiments and opinions will be made~\cite{abdar2021review}.\\
Fig.~\ref{Fig: Images} shows the predictive posterior distribution for two non-cancer (normal) images obtained by the MCD method (200 MC iterations). Here the predictive entropy is used as the uncertainty tool; the closer it is to zero, the higher the value of confidence, and the case should be reported as normal. For the case shown in Fig.~\ref{fig: Certain}, the anticipated output is normal, and the model is confident about its prediction (a small amount of predictive entropy of 0.12). On the contrary, for some scenarios, the model estimate is incorrect for the right diagram in Fig.~\ref{fig: Uncertain} (the model will predict this image randomly). Nevertheless, the model calculates a high amount of predictive entropy of 0.69. As a result of this, it leads to the lack of confidence in this particular prediction and declares that "I am not sure." Since the model is not assured about its forecast, the image could be sent to an expert for an extra opinion.

\section{Method}\label{Sec:Method}
The importance of knowing how much a model is confident about its prediction is inevitable, especially in the field of medical and autonomous vehicles. In analyzing the forecasts, sometimes it is required to be able to calculate the certainty and tell that “perhaps the output prediction will be better with more data? or “change the model?” or  “be careful in your decision.”. Unfortunately, most deep learning models cannot answer these questions and are viewed as deterministic models and consequently noticed as functioning differently from the probabilistic models that own uncertainty information. The conditions that can result in uncertainty consist of~\cite{hullermeier2021aleatoric}:
\begin{itemize}
    \item Noisy data, which will lead to aleatoric uncertainty.  It is due to the natural stochasticity of observations. This kind of uncertainty is complicated and irreducible. It brings up the concept of randomness. The exemplary specimen of aleatoric uncertainty is the coin flipping: The data-generation process in this kind of experiment has a stochastic element that cannot be condensed by any additional source of data (except for Laplace’s demon). As a result, even the finest model of this procedure will only offer probabilities for the two probable outcomes, tails and heads, but no definite answer. Fig.~\ref{fig: slide} represents a sinus diagram that was sampled around the left cloud and the right cloud.  According to Fig.~\ref{fig: slide} in the left cloud of the diagram the most data scatters around the main function, shows more aleatoric uncertainty, which might cause by different kinds of noise, such as sensor malfunction. Extra measurements cannot reduce this uncertainty until the noise source is removed; for example, the sensor keeps generating errors around the left cloud for the sensor malfunctioning example. On the contrary, we can see that around the right cloud, the training data fits the main function, which shows low aleatoric uncertainty.
    
    \item Uncertainty in model parameters, which will lead to epistemic uncertainty. It is arising from limited data and knowledge. In other words, it refers to ignorance. Contrary to the aleatoric uncertainty, this kind of uncertainty is reducible by using a diverse dataset or changing the model. Epistemic uncertainty, as mentioned, can rise in parts in which there is a smaller number of samples for training. This is the situation of the left, middle and right parts of our clouds in Fig.~\ref{fig: slide}. By feeding additional data in those spaces, uncertainty would decrease. In high-risk and hazardous applications, it is essential to recognize such areas.
    
\end{itemize}

Bayesian statistics can allow working on the weights’ distribution instead of fixed weights (point estimates), which will solve the problem of epistemic uncertainty.\\
In this paper, we concentrate on epistemic uncertainty. The tool to calculate uncertainty is entropy which will be discussed in detail.

\begin{figure*}[!t]
\centering
\begin{minipage}[b]{.27\textwidth}
  \subfloat[EMCD]{\includegraphics[width=\textwidth]{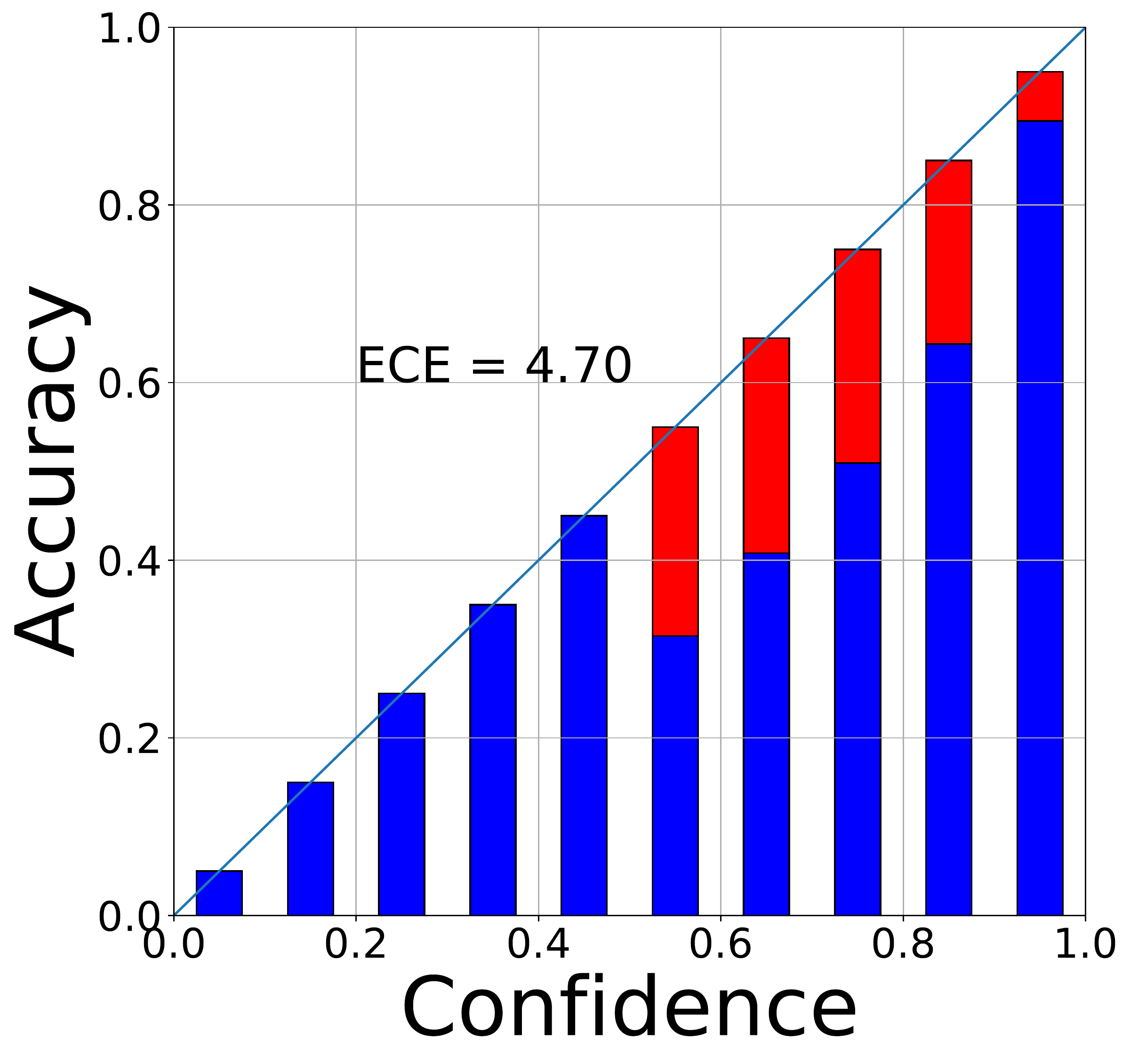}}
  \label{Fig:ECE of MCD for CT}
\end{minipage}\qquad
\begin{minipage}[b]{.27\textwidth}
  \subfloat[MCD]{\includegraphics[width=\textwidth]{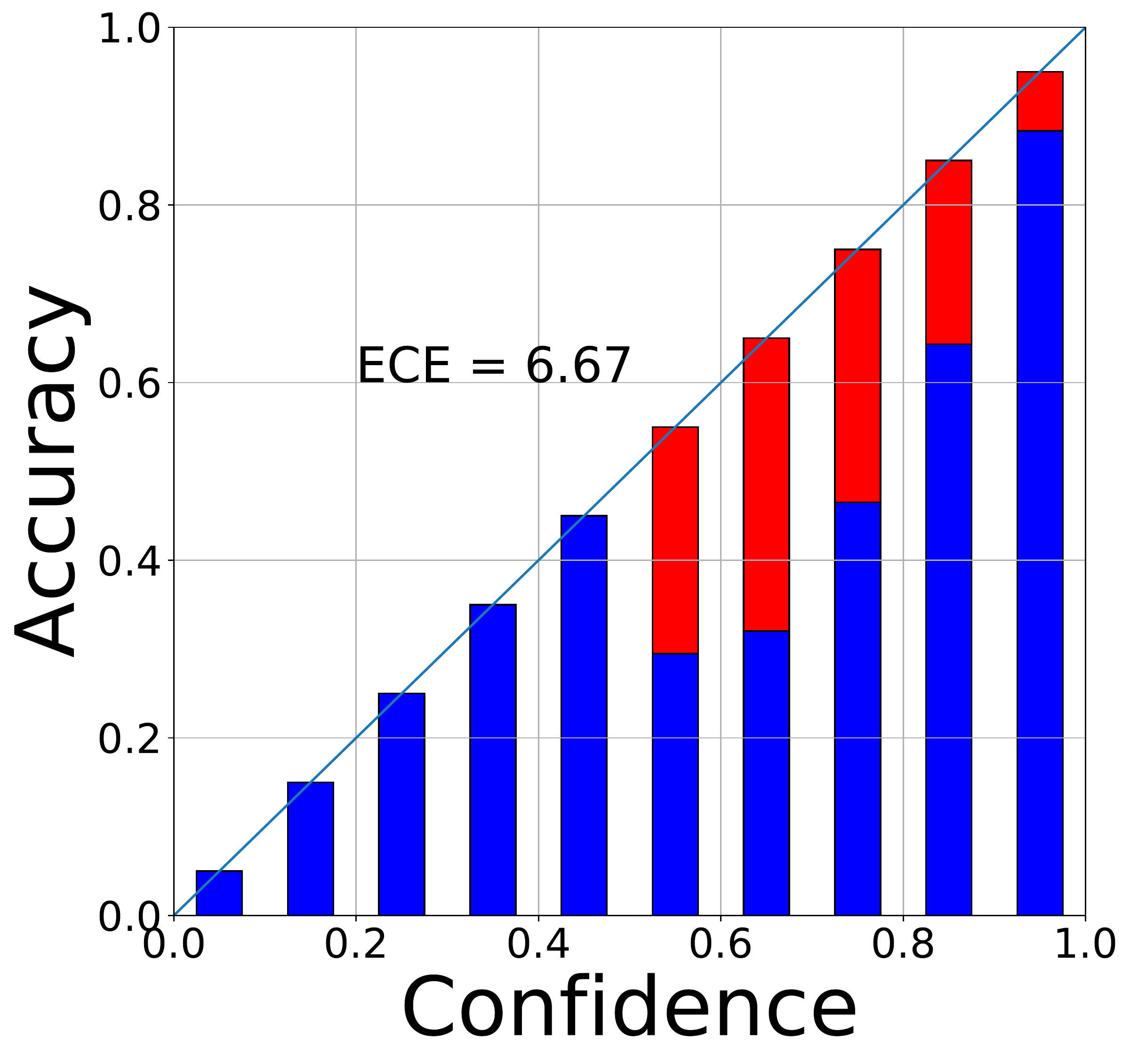}}
  \label{Fig:ECE of EMCD for CT}
\end{minipage}\qquad
\begin{minipage}[b]{.27\textwidth}
  \subfloat[Ensemble]{\includegraphics[width=\textwidth]{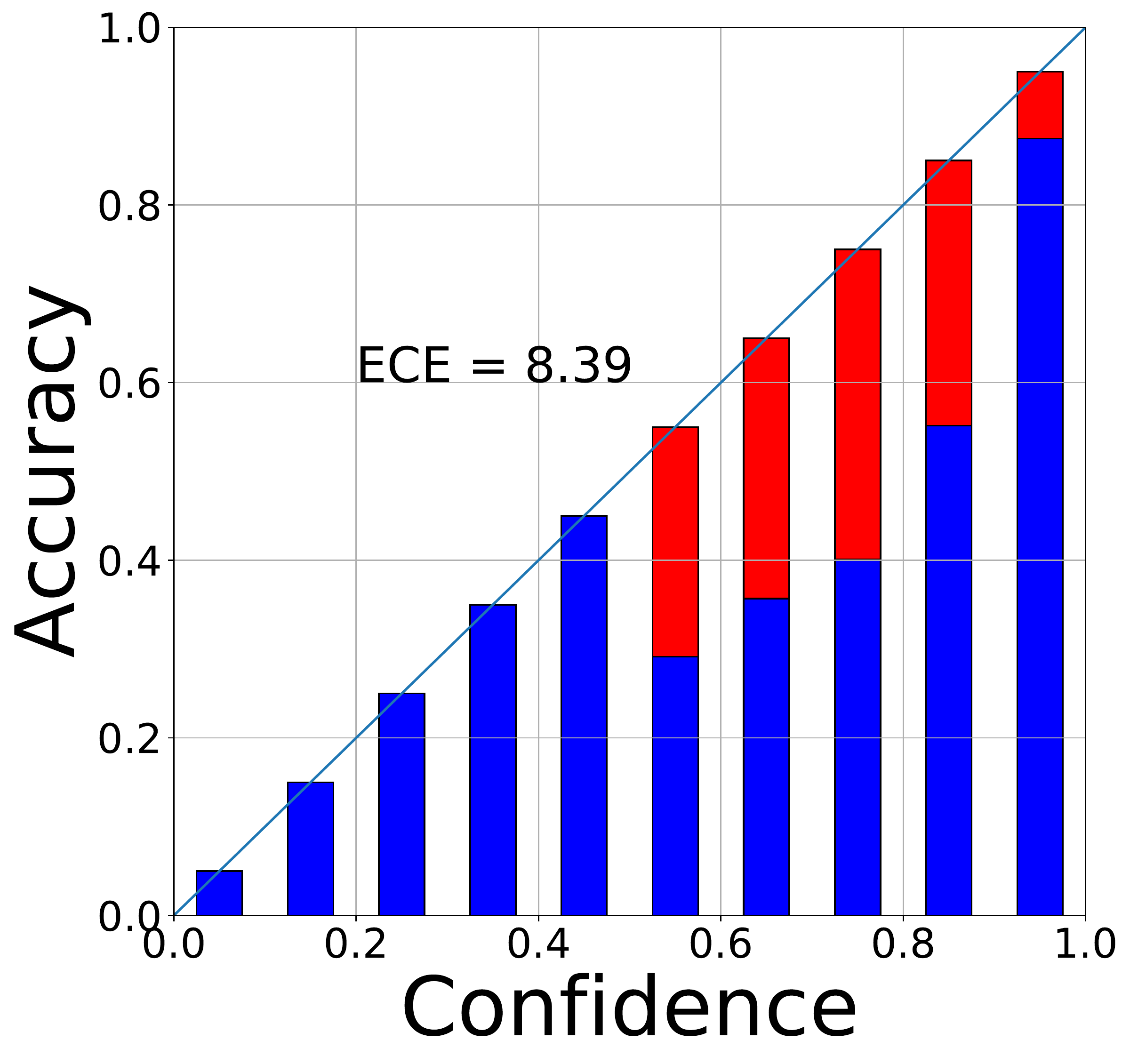}}
  \label{Fig:ECE of Ensemble for CT}
\end{minipage}
    \caption{The ECE plots (reliabilty diagram) are depicted.}

    \label{Fig:final ECE}
\end{figure*}

\subsection{MCD}
 Gal proposed that the posterior distribution of a bayesian setting can be estimated by enabling dropout not only in the training phase but also during the testing phase~\cite{gal2016dropout}.In this method, for the test time, multiple estimations of a specific test sample will be generated and then used to calculate the average of them as the final posterior.
It is essential to note that MCD is a Gaussian Process (GP) approximation with three hyperparameters: retaining rate, model precision, and length scale. More information about MCD can be found are available in~\cite{polikar2012ensemble,seoh2020qualitative}.
For a typical test input, the predictive mean ($\mu_{pred}$) of this model is approximated as below:
\begin{equation}
    \mu_{pred} \approx \frac{1}{T} \sum_t p(y = c | x, \hat\omega_t)
\end{equation}
\noindent where $x$ is the test input, $p(y = c | x, \hat\omega_t)$ is the probability that $y$ belongs to $c$ which is the output of softmax, and $w_{th}$ is the set of parameters of the $t$ forward pass. $T$ is the number of MC iterations (forward passes). The variance of the ultimate distribution is also known as the predictive entropy (PE). Following~\cite{gal2016dropout}, the predictive entropy (PE), can be treated as the uncertainty estimate generated by the trained model:
\begin{equation}\label{Eq:MC-Dropout-PE}
    PE = - \sum_c \mu_{pred} \log \mu_{pred}
\end{equation}
where $c$ ranges over the classes, which in this work is two. The smaller the PE, the more confident the model about its prediction.


\subsection{Ensemble Bayesian Networks}

We are using an ensemble to benefit from the strength of combinations of networks. Generally, there are two methods for developing an ensemble; ensembling on the data and ensembling on the model\cite{}. In this work, we use the latter one. To get the ideal result of using model ensembling, the networks must be different far from each other. In most of the previous works, the authors concentrate on random parameter initialization to defferentiate DNNs. In this work, in addition, to use this technique, we vary the number of hidden layers in each layer accordingly, the networks are sufficiently different and the correlation between the results will be reduced to a near-realistic average.  Each network will predict a probability, and the mean of those probabilities will be the ultimate predictive entropy (posterior). The PE measure in this situation will be calculated as follow~\cite{van2020simple}:

\begin{equation}
    \hat{p} (y|x)\ = \ \frac{1}{N}\ \sum\limits_{i=1}^N p_{\theta_{i}} (y|x) \label{eq:1}
\end{equation}

\begin{equation}
   PE \ = \ \sum\limits_{i=0}^C \hat{p}(y_i | x)\ log\ \hat{p}(y_i | x) \label{eq:2}
\end{equation}

\noindent where $\theta_i$ represents the set of parameters of $i_{th}$ network element, and $C$ ranges over two classes.

\subsection{EMCD}
A combination of the MCD algorithm and ensemble networks will produce EMCD. Through this, we can use the advantages of both algorithms proposed in the previous subsections~\cite{kendall2017uncertainties} 
The ensemble network here consists of different DNNs with different architectures. By performing several stochastic forward passes, the evaluation of each network will be done by MCD. Finally, a single Gaussian distribution will be calculated by averaging all posteriors probabilities. The PE calculation is similar to the ensemble but different in the calculation of the posteriors:

\begin{equation}
    \hat{p} (y|x) \approx \frac{1}{T} \sum_{t=1}^{T} p(\hat{y} | \hat{x},\hat\omega_t )
\end{equation}

\begin{equation}
   PE \ = \ \sum\limits_{i=0}^C \hat{p}(y_i | x)\ log\ \hat{p}(y_i | x) \label{eq:2}
\end{equation}

\noindent where $\hat\omega_t$ are the parameters of the model and $C$ ranges over two classes.

\section{Evaluation of Uncertainty estimates}\label{Sec:Uncertainty quantification}

Here quantitative performance evaluation metrics for predictive uncertainty estimates will be introduced. Contrary to~\cite{ghoshal2020estimating}, the purpose is to do a quantitative and objective evaluation of the predictive uncertainty estimates. Estimates are first compared with ground truth labels and categorized into two groups: correct and incorrect. Predictive uncertainty estimates are similarly compared with a threshold and put into two groups: certain and uncertain. The mixture of correctness and confidence groups results in four possible outcomes as shown in Fig.~\ref{Fig: Confusion matrix}: (i) correct and certain point to true certainty (TC), (ii) incorrect and uncertain indicate true uncertainty (TU), correct and uncertain indicated by false uncertainty (FU), and (iv) incorrect and certain indicate false certainty (FC). TC and TU are the diagonal and favorite results. FU is known as the fortunate outcome as an uncertain prediction is true. Finally, FC is the worst outcome as the network has confidently made an incorrect prediction. Based on these outcomes, multiple quantitative performance metrics to objectively quantify predictive uncertainty estimates will be introduced:
\begin{itemize}
    \item Uncertainty sensitivity (USen): USen is calculated as the number of false and uncertain estimates divided by the overall number of false estimates: 

    \begin{equation}\label{Eq:USen}
        USen \ = \ \frac{TU}{TU + FC}
    \end{equation}

USen or uncertainty recall (URec) is related to the sensitivity (recall) or true positive amount of the conventional confusion matrix.
 USen has a high priority because it evaluates the power of the model to transfer its assurance in misclassified examples.
 
 \item Uncertainty Specificity (USpe): USpe measures as the number of correct and certain predictions (TC) divided by the overall number of correct predictions:
 
    \begin{equation}\label{Eq:USpe}
        USpe \ = \ \frac{TC}{TC + FU}
    \end{equation}
    
USpe or correct certain ratio is comparable to the specificity performance metric.

\item Uncertainty precision (UPre): UPre is calculated as the number of false and uncertain predictions divided by the overall number of uncertain predictions: 

    \begin{equation}\label{Eq:UPre}
        UPre \ = \ \frac{TU}{TU + FU}
    \end{equation}

UPre has the identical concept of precision in the traditional binary classification. 

\item Uncertainty accuracy (UAcc): Analogous to the accuracy of classifiers, the UAcc is calculated as the number of total diagonal results divided by the overall number of outcomes:

    \begin{equation}\label{Eq:UAcc}
        UAcc \ = \ \frac{TU + TC}{TU + TC + FU + FC}
    \end{equation}
    
\end{itemize}

\begin{figure}
	\centering
	
		\includegraphics[width=\linewidth]{ 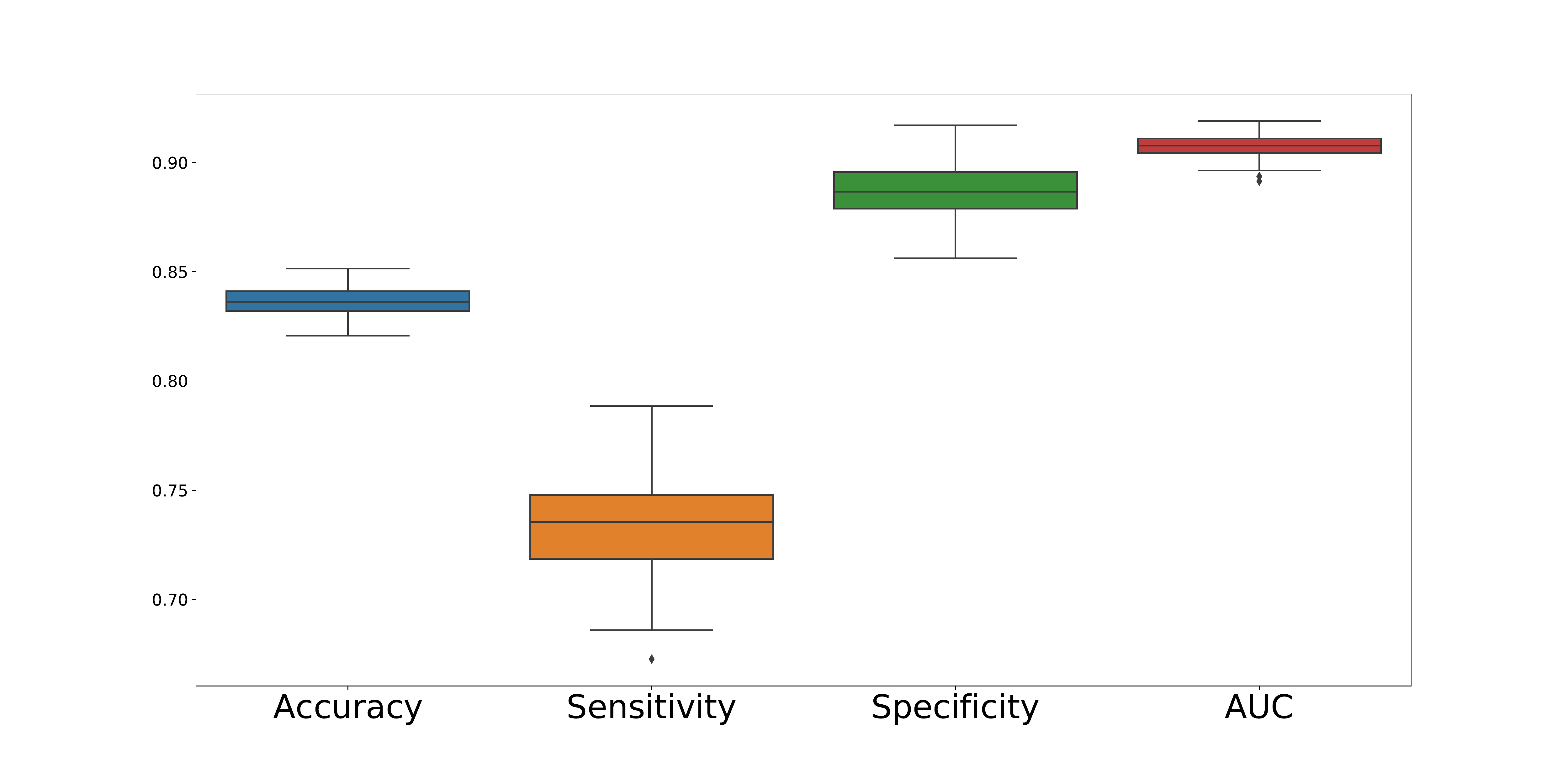}
		\caption{The box plot of accuracy, sensitivity, specificity, and AUC for the base model by performing $100$ individual runs.}
		\label{fig: BoxPlot}
		
     \end{figure}

\begin{table}[!t]
\centering
\caption{Accuracy, sensitivity, specificity, and AUC values for mean and std by performing $100$ individual runs for the base model. }\label{Tab: mean & std }
\begin{tabular}{llllll}
        \hline
        Performance Metric           & Mean   & std  \\
        \hline
        Accuracy          & 0.836142 & 0.006557     \\
        Sensitivity         & 0.734022 & 0.024095    \\
       Specificity	      & 0.886593 & 0.012117 \\
       AUC	      & 0.907571& 0.005323   \\
        \hline
    \end{tabular}
\end{table}
\begin{figure*}[!h]
\centering
\begin{minipage}[b]{.27\textwidth}
  \subfloat[MCD]{\includegraphics[width=\textwidth]{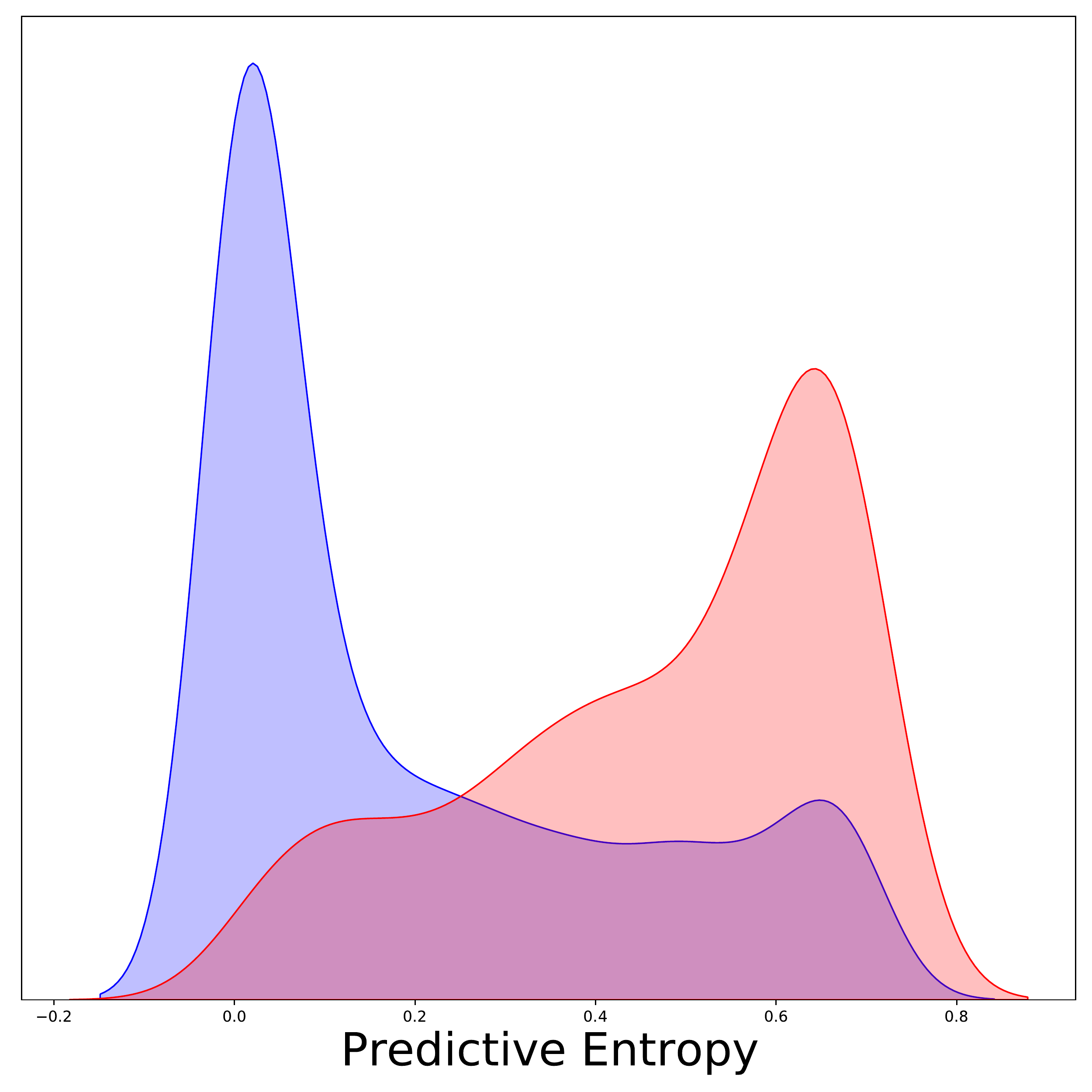}}
  \label{Fig: PE MCD}
\end{minipage}\qquad
\begin{minipage}[b]{.27\textwidth}
  \subfloat[EMCD]{\includegraphics[width=\textwidth]{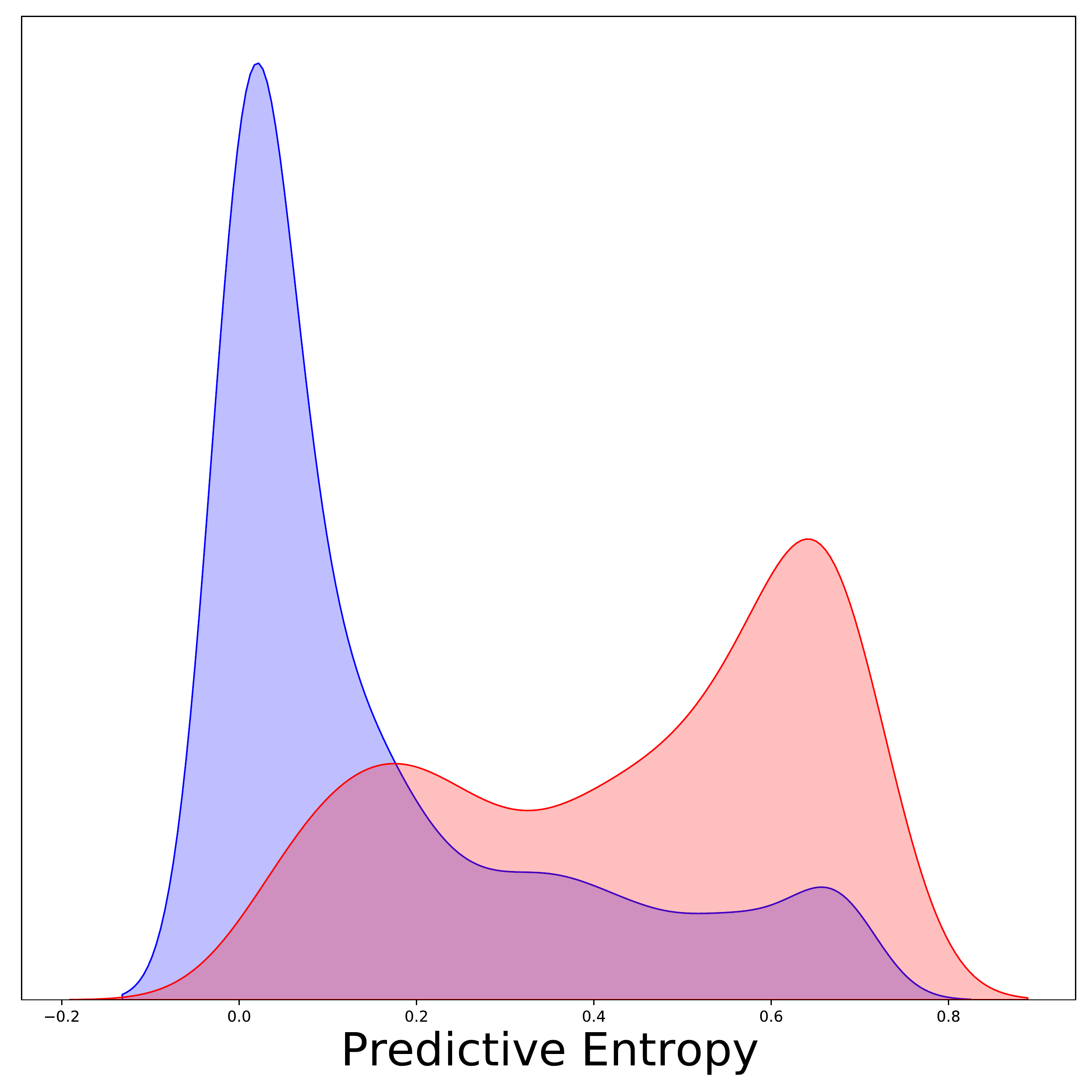}}
  \label{Fig:PE EMCD}
\end{minipage}\qquad
\begin{minipage}[b]{.27\textwidth}
  \subfloat[Ensemble]{\includegraphics[width=\textwidth]{HMnist_28_28_RGB_MCDropout_Ensemble_Entropy.pdf}}
  \label{Fig:PE Ensemble}
\end{minipage}
    \caption{The distributions of correctly and miss classified test outputs sorted by predicted entropy}

    \label{Fig:PE}
\end{figure*}
\begin{figure*}
\centering
\begin{minipage}[b]{.3\textwidth}
    \subfloat[Unc. accuracy]{\includegraphics[width=\textwidth]{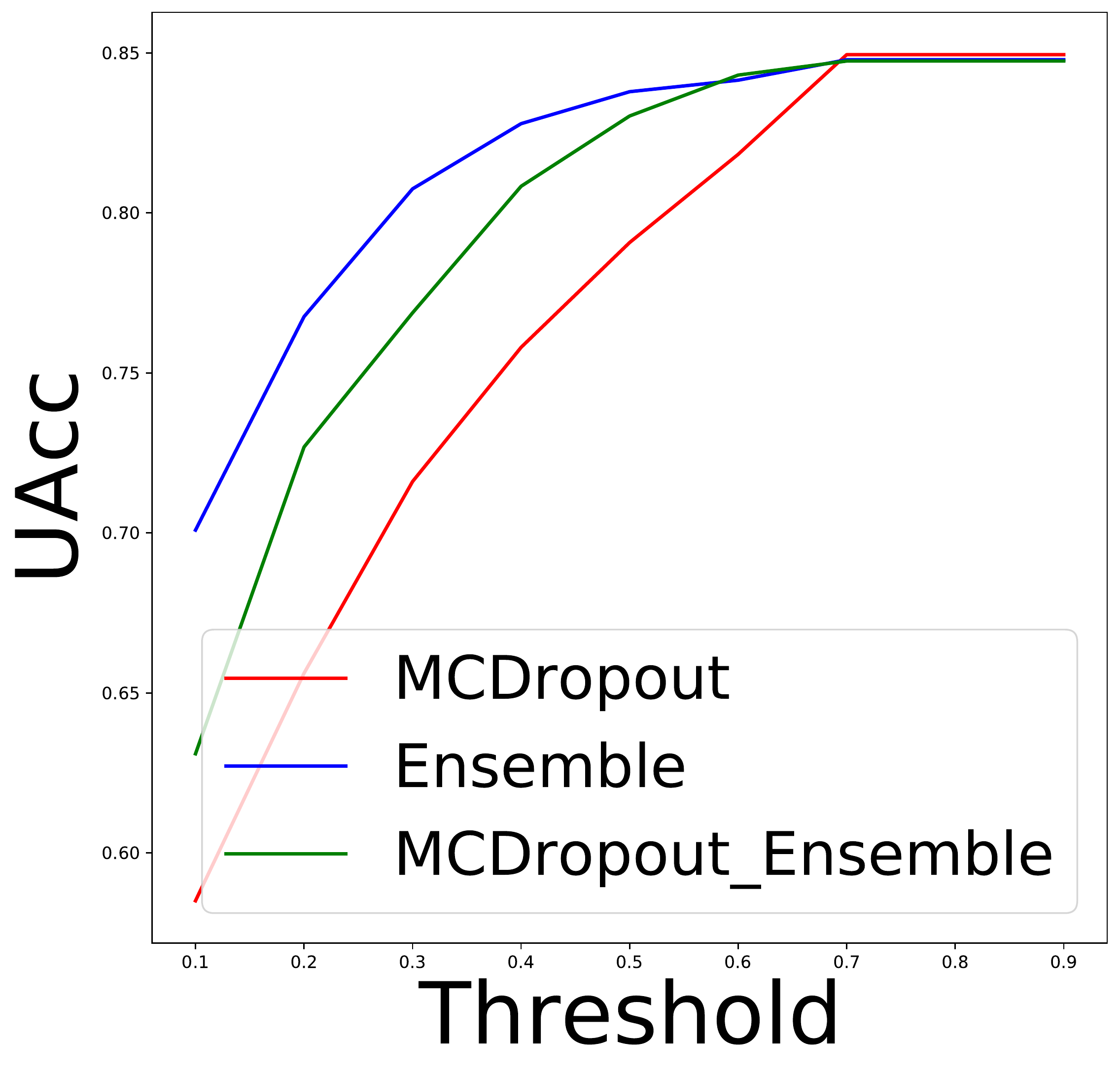}}
    \label{Fig:UAcc}
\end{minipage}\qquad
\begin{minipage}[b]{.3\textwidth}
    \subfloat[Unc. sensitivity]{\includegraphics[width=\textwidth]{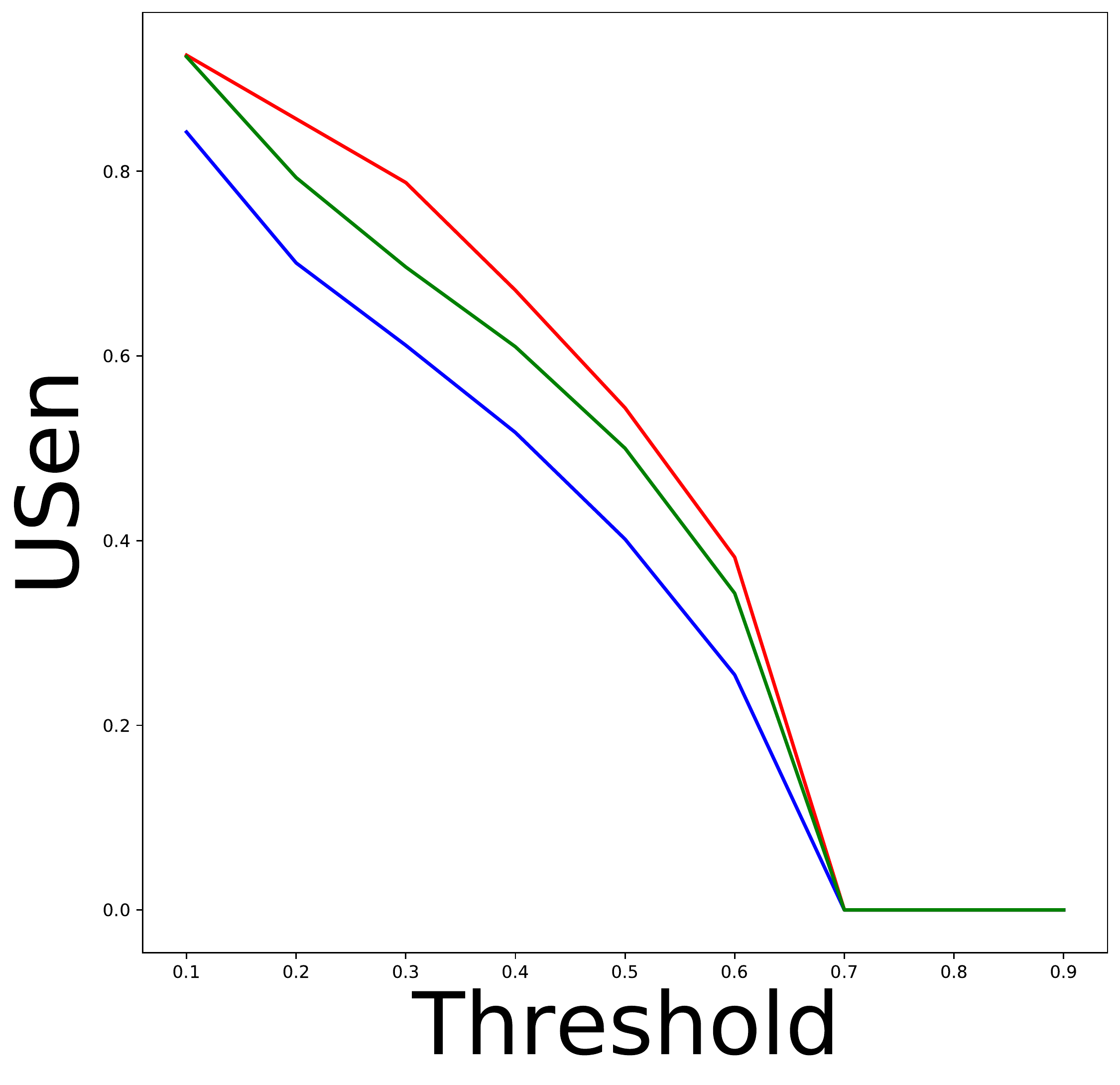}}
    \label{Fig:USen}
\end{minipage}\qquad
\begin{minipage}[b]{.3\textwidth}
    \subfloat[Unc. specificity]{\includegraphics[width=\textwidth]{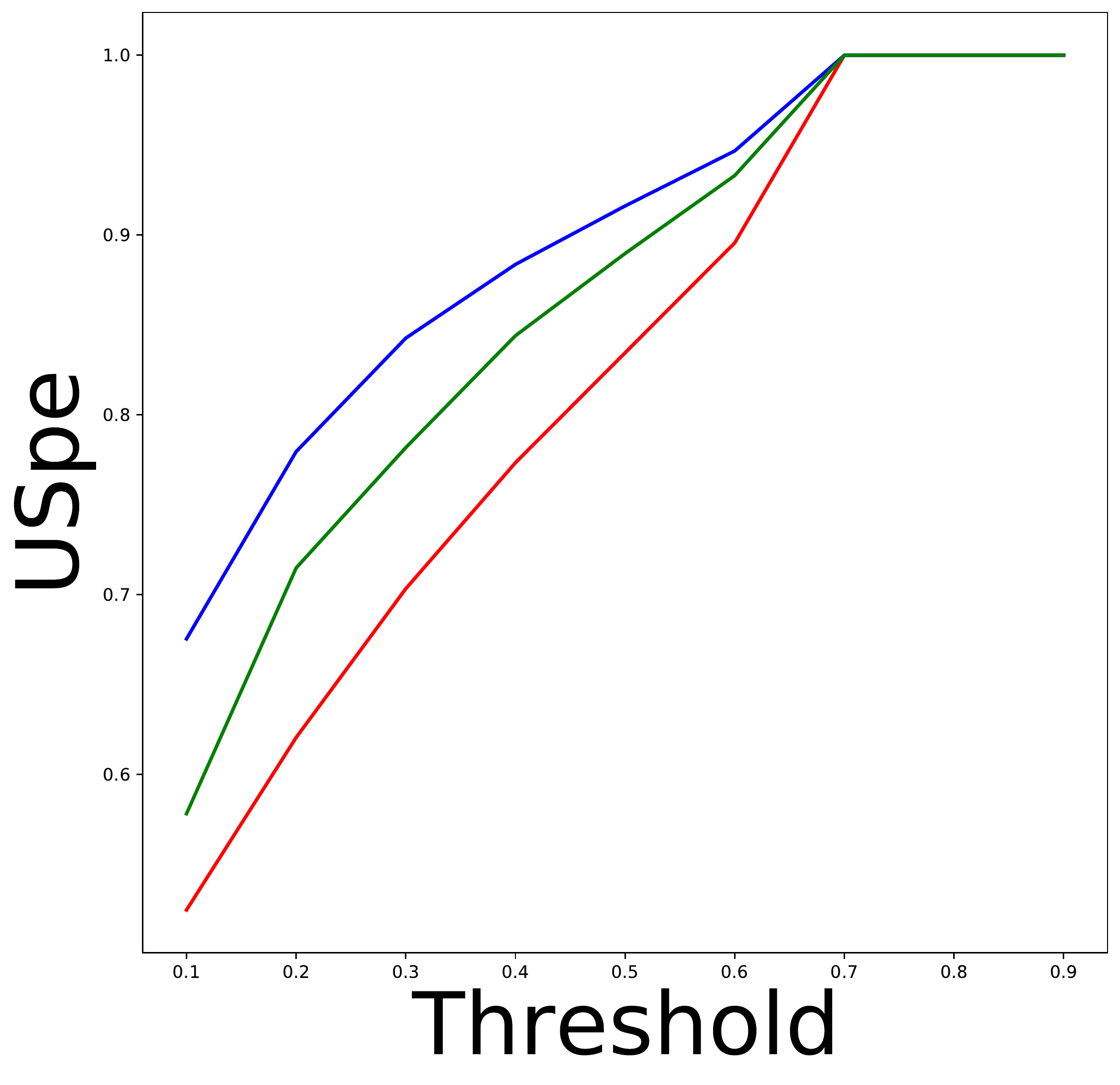}}
    \label{Fig:USpe}
\end{minipage}\qquad
\begin{minipage}[b]{.3\textwidth}
    \subfloat[Unc. precision]{\includegraphics[width=\textwidth]{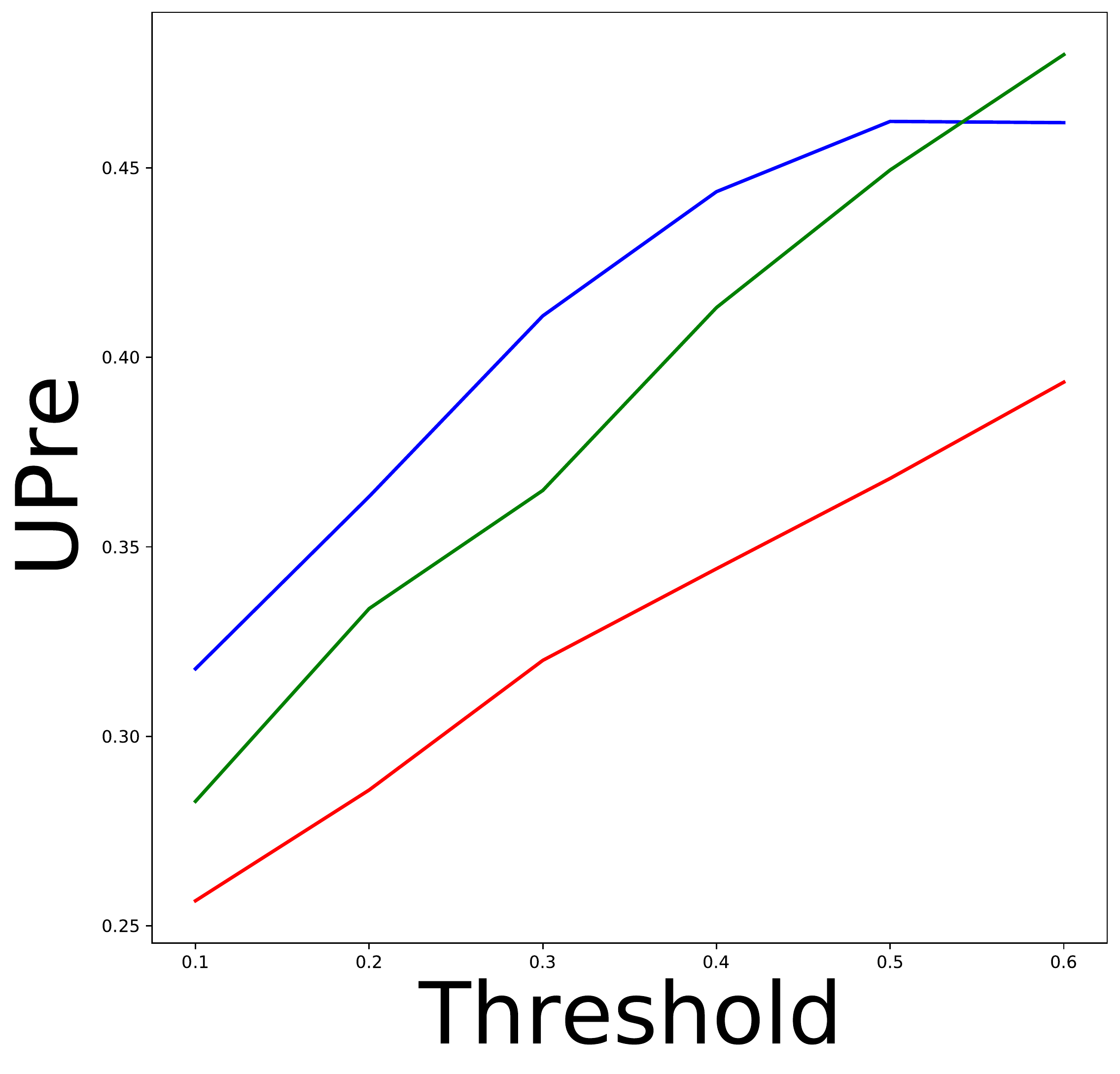}}
    \label{Fig:UPre}
\end{minipage}
    \caption{Three uncertainty quantification techniques are compared quantitavely for different thresholds}
    \label{Fig:UncMetrics}
\end{figure*}
An reliable model will produce a high UAcc. The best amount of USen, USpe, UPre, and UAcc are one, while the worst are zero. It is required to have these metrics as close as possible to one to have a suitable model. Obtaining USen, USpe, and UPre close to one means that the network has self-awareness about what it knows and what it does not know. Such a network can tell us when a user can trust network’s predictions, because the network confidently measures and declares its confidence (for example, in estimating predicted uncertainty).

\section{Experimental Setup}\label{Sec:Experimental}

To compensate for the lack of data to train the network from scratch, transfer learning is used, trained on the ImageNet dataset. The dataset is divided to $80\%$ and $20\%$ between training and testing subsets. The images in the dataset first are standardized and resized to 224*224 before entering the network.
This concludes in 50176 convolutional features, which afterward will be processed by fully connected layers with softmax. Relu is used as the activation function in this network, and the dropout rate of 0.25 is used for model development using three uncertainty quantification methods. The Adam algorithm with a learning rate of 0.001 is used for the cross-entropy loss function optimization. The number of neurons for the MCD model in three fully connected layers are set to 512, 256, and 64 correspondingly. The ensemble model comprises 30 separate networks in which the hidden layers are randomly selected between two and three. Similarly, the number of neurons in fully connected layers are randomly picked between (512, 1024), (128, 512), (8, 128), respectively. The only difference between EMCD and ensemble is that in EMCD, the evaluation of each network is done by the MCD algorithm.

\section{Result Analysis and Discussion}\label{Sec:Results}

For gaining insights into our models, we analyze the performance metrics of our base model (similar to MCD without activating dropout at test time which results in a simple non-Bayesian algorithm). The base model is trained 100 times individually, and the performance metrics are calculated for each run, and formerly, the box plot graph is plotted. Fig. ~\ref{fig: BoxPlot} displays the box plots for the skin cancer dataset, respectively, for accuracy, sensitivity, specificity, and AUC. It should be noted that PCA is not used for the calculation of these values, and all features are passed to the classifier. To systematically compare different architectures used in this paper and achieve a reliable estimate of the sample label, for mean and standard deviation of all predictions, metrics, as mentioned above, are calculated and reported in table ~\ref{Tab: mean & std }.


Before the uncertainty estimation evaluation, first, the calibration of predictions generated by DNNs are checked. Fig.~\ref{Fig:final ECE} shows the expected calibration error (ECE) for three uncertainty quantification methods. ECE demonstrates the model accuracy as a function of confidence~\cite{guo2017calibration}. To assess ECE, predictions are clustered in different bins (here M bins) based on their confidence (the value of the max softmax output). The calibration error of each bin calculates the variance between the fraction of correctly classified predictions (accuracy) and the average of the probabilities (confidence). ECE is a weighted average of this error through all bins~\cite{guo2017calibration}:
\begin{equation}\label{Eq:ECE}
    ECE = \sum_{m=1}^{M} \frac{|B_m|}{n} \left | acc(B_m) - conf(B_m) \right |
\end{equation}

In Fig.~\ref{Fig:final ECE}, the more the blue portion deviates from the equivalent red parts, the less calibrated the NN model. A completely calibrated model will produce the identity line in this diagram. The current plot clearly shows that probabilities generated by three DNN families examined in this paper are not calibrated. Based on the ECE values stated in Fig.~\ref{Fig:PE}, all models classify samples with too much confidence, which leads to deceptive results. EMCD has the minimum ECE value of 4.70. \\
We then examine the model overall ability to measure and report its uncertainty predictions. Fig.~\ref{Fig:PE} shows the marginal distributions of uncertainty for correct (blue) and incorrect (red) predictions. The plots display that the centers of two groups are apart from each other (though there may be some long legs). The visual inspection of the three sub-studies shows that, on average, incorrect predictions have high uncertainty. This qualitative research highlights the model ability to measure and transfer confidence (or uncertainty) in generated predictions. This finding is of great practical importance because reliable uncertainty estimates provide additional valuable information to the predicted probabilities. These can be used to flag uncertain predictions and request a second opinion from a medical professional.
 We also comprehensively assess the predictive uncertainty using performance metrics introduced in Section ~\ref{Sec:Experimental}. Fig.\ref{Fig:UncMetrics} shows uncertainty accuracy (UAcc), uncertainty sensitivity (USen), uncertainty specificity (USpe), and uncertainty.
 Precision (UPre) values are calculated for uncertainty thresholds in the range of 0.1 to 0.9. UAcc, USpe, and UPre are positively correlated with the uncertainty threshold. This correlation is negative for USen. UAcc, USen, and USpe all obtain values close to one for a wider range of thresholds. Obtaining a high USen means that all three uncertainty quantification techniques are capable of flagging incorrect predictions with high uncertainty. These methods are quite clever in flagging wrong predictions for additional examination. However, none of the uncertainty quantification methods accomplishes a high UPre close to one. This is because many correct predictions have a high uncertainty (FU). This can be seen in the long tail of the predicted uncertainty distribution in Figure~\ref{Fig:PE}. It is also noteworthy that correctly classified images are much larger than misclassified images. This leads to a smaller number of TU predictions than FU, causing a low Upre. This is an anticipated pattern for models with high accuracy.
 \begin{table}[!t]
\centering
\caption{The performance of each algorithm for treshold$= 0.4$}\label{Tab:UncMetrics}
\begin{tabular}{llllll}
        \hline
        UQ Method           & UAcc   & USen  & USpe   & UPre \\
        \hline
        MCD          & 76\% & 0.671  & 0.77   & 0.344    \\
        EMCD         & 83\% & 0.58  & 0.83   & 0.41    \\
        Ensemble     & 84.8\% & 0.52  & 0.88   & 0.44   \\
        \hline
    \end{tabular}
\end{table}
 
Choosing the best amount of uncertainty threshold depends on users' preferences. The suitable threshold is between 0.3 and 0.7. In Table~\ref{Tab:UncMetrics}, the uncertainty evaluation metrics are reported for all three methods for a threshold equal to 0.4. UAcc for the MCD method smaller than the other two ensemble methods, which is 76\%. The ensemble algorithm reaches the maximum UAcc among uncertainty quantification algorithms, and its UAcc (84.8\%) is to some extent larger than that of EMCD (83\%). There is a similar pattern for other performance metrics for the three methods.

\section{Conclusion}\label{Sec:Concl}
This paper investigates the capability of deep uncertainty quantification methods for skin cancer detection. A newly proposed confusion matrix and various performance metrics for assessing predictive uncertainty estimates are used on three algorithms. Through comprehensive assessment, we find and prove that ensemble techniques have better functionality related to their predictions, resulting in more reliable diagnosis solutions. There is a large scope for improving the uncertainty evaluation metrics and their application for DNN improvement. For future work, one can optimize each algorithm to improve the UAcc and other parameters related to the confusion matrix of uncertainty.
This will result in optimized networks based on performance metrics for both point predictions and uncertainty estimates.


\end{document}